\documentstyle[12pt]{article}
\begin{document}
\newcommand{\bea}{\begin{eqnarray}}
\newcommand{\eea}{\end{eqnarray}}
\newcommand{\be}{\begin{equation}}
\newcommand{\ee}{\end{equation}}
\newcommand{\non}{\nonumber}
\newcommand{\ov}{\overline}
\global\parskip 6pt
\begin{titlepage}
\begin{center}
{\Large\bf Homology in Abelian Lattice Models}\\
\vskip 1.0in
Mark Rakowski \footnote{Email: Rakowski@maths.tcd.ie}   \\
\vskip .10in
{\em Dublin Institute for Advanced Studies, 10 Burlington Road, }\\
{\em Dublin 4, Ireland}\\
\vskip .10in
and\\
\vskip .10in
Siddhartha Sen \\
\vskip .10in
{\em School of Mathematics, Trinity College, Dublin 2, Ireland}
\end{center}
\vskip .10in
\begin{abstract}
   We study abelian lattice gauge theory defined on a simplicial complex
with arbitrary topology. The use of dual
objects allows one to reformulate the theory in terms of different
dynamical variables; however, we avoid the use of the
dual cell complex entirely. Topological modes which are present in the
transformation now appear as homology classes, in contrast to the
cohomology modes found in the dual cell picture. Irregularities
of dual cell complexes do not arise in this approach. We treat
the two and three dimensional cases in detail.
\end{abstract}
\vskip 1.5in
\begin{center}
PACS numbers: 11.15.Ha, 05.50.+q
\end{center}

\end{titlepage}

\section{Introduction}

   Duality transformations have been studied in statistical systems
since the early work of Kramers and Wannier \cite{KW}. It is a
striking result that the high and low temperature properties of some
theories are related by this method. There have been many other
applications of this idea especially in the case of hypercubic
lattices; see \cite{RS} for an extensive review. Looking at duality
from the more general framework of a simplicial complex, one finds that
topological modes generically enter into these duality transformations
\cite{DRUHL,RAK}. Even the two dimensional Ising model on a simple
square lattice (torus) has such modes.

   While theories defined on hypercubic lattices are particularly simple
to study since those lattices are self-dual, one may be interested in
approximating continuum models defined on spaces with different
topology. Unfortunately, the usual notion of the dual cell complex
\cite{JM} associated to a given simplicial complex is in general
very irregular in spite of the regularity of simplicial complexes;
duality may map simplicial objects into polygons of various type. In
\cite{RAK}, it was shown that dual theories, defined on dual cell
complexes, generally have topological modes which are in correspondence
with cohomology classes on the dual cell complex.

   In this paper, we
point out that duality can be considered without reference to
dual cell complexes, and the cohomological modes which previously
entered the analysis will now appear as homology classes on the
original simplicial complex. While duality changes the nature of
the dynamical variables in a given theory, we will here use only
the original simplicial complex in formulating the dual theory.

   In the following section we review the essentials of simplicial
complexes and homology. Applications of our approach to duality
will be illustrated in $d$-dimensional abelian gauge theory. The
two dimensional case will then be fully analyzed on a general
combinatorial 2-manifold and the partition function reduced to
a single mode sum for any plaquette based action (not necessarily
ones which are subdivision invariant). Finally, we treat the
three dimensional case in some detail.

\section{Simplicial Complexes and Homology}

   Let us begin by recalling some standard material on simplicial
complexes and homology; we refer to \cite{JM} for a complete treatment.

   Intuitively, a simplicial complex is a collection of
simplices of various dimensions (points, line segments, triangles, etc.)
which are glued together in a regular way. More formally,
let $V=\{ v_{1},...,v_{N_{0}} \}$ be a collection of $N_{0}$
elements which we will call the {\em vertex} {\em set}. A {\em
simplicial complex} $K$ is a collection of finite nonempty subsets
of $V$ such that if $\sigma \in K$ so is every nonempty subset of
$\sigma$. An element of $K$ is called a {\em simplex} and its {\em
dimension} is one less than the number of vertices it contains. We
picture the 1-dimensional simplex $\{ v_{i}, v_{j} \}$ as the line
segment connecting two distinct points in Euclidean space.
Similarly, $\{ v_{i},v_{j},v_{k} \}$ can be pictured as the triangle
with the three indicated vertices. An {\em orientation} of a simplex
$\{ v_{0},...,v_{m} \}$, denoted $[ v_{0},...,v_{m} ]$, is an
equivalence class of the ordering of the vertices according to even
and odd permutations. This gives direction to a line segment and
to the circulation around the boundary of a triangle, and so
on for higher dimensional simplices.

   Let $K^{(m)}=\{ \sigma_{\alpha} \}$ denote the collection of all oriented
$m$-simplices in $K$. The group of $m-chains$ on $K$ with coefficients
in an abelian group $G$, denoted by
$C_{m}(K,G)$,
is defined to be the set of all finite linear combinations $\sum_{\alpha}
\, n_{\alpha}\,\sigma_{\alpha}$, $n_{\alpha}\in G$, with the natural
componentwise addition of chains as the group operation.
A {\em boundary}
operator $\partial_{m}: C_{m}(K,G)\rightarrow C_{m-1}(K,G)$ is defined
on a given simplex
\be
\partial_{m} [v_{0},...,v_{m}] = \sum_{i=0}^{m} \; (-1)^{i} \;
[v_{0},...,\bar{v}_{i},...,v_{m}] \;\; ,
\ee
(where we have omitted the vertex corresponding to $\bar{v}_{i}$)
and then extended to all of $C_{m}(K,G)$ by linearity.
Let $Z_{m}(K,G)$ denote the kernel of $\partial_{m}$ and $B_{m}(K,G)$
the image of $\partial_{m+1}$; the {\em homology group} $H_{m}(K,G)$
is then defined as the quotient group $Z_{m}/B_{m}$. $Z_{m}(K,G)$
is called the group of $m$-cycles and $B_{m}(K,G)$ the group
of $m$-boundaries. It is a
nontrivial theorem \cite{JM} that the homology groups are topological
invariants, and hence independent of subdivision of the complex.
We will have no need of either cohomology or dual block (cell) complexes
in this paper.

   Applications to physical systems naturally focus on discrete
approximations to smooth manifolds. One fact we will use is that a
simplicial complex $K$ which models a smooth manifold $M$ of
dimension $d$ without boundary,
consists of $d$-simplices which are glued pairwise
along faces of dimension $d-1$, so all $(d-1)$-simplices in $K$ are
common to precisely two $d$-simplices. For example, we can represent
the three dimensional sphere $S^{3}$ as the boundary of a single
$4$-simplex,
\bea
\partial\, [0,1,2,3,4] = [1,2,3,4] - [0,2,3,4] + [0,1,3,4] - [0,1,2,4]
+ [0,1,2,3] \;\; .\label{cs3}
\eea
In this case the complex $K$ consists of the above $3$-simplices
together with all their subsimplices. Note, for example, that the
$2$-simplex
$[2,3,4]$ is common to only the first two $3$-simplices in the
list, in accordance with the pairwise gluing condition.

\section{Duality in Gauge Theory}

  We begin our analysis with the case of $Z_{P}$ lattice gauge theory
on a closed $d$ dimensional simplicial complex $K$. By definition, such a
theory is specified by an action $S$ which is a function of the link
variables $U_{ij}$ only through the holonomy
\bea
U_{[i,j,k]} = U_{ij} U_{jk} U_{ki} \;\; .
\eea
The partition function of $Z_{P}$ lattice gauge theory is defined as
a sum over all link variables $ U_{ij}\in Z_{P} $, which we represent
multiplicatively, and a Boltzmann weight factor for every 2-simplex
in the simplicial complex $K$ \cite{MC},
\be
{\cal Z}_{d} = P^{-N_{1}}\; \sum_{\{ U_{ij} \} } \;
\prod_{ \Delta \in K^{(2)} } \; \exp [ S(U_{\Delta}) ] \;\;
.\label{z}
\ee
The factor of $P^{-N_{1}}$, where $N_{1}$ is the number of $1$-simplices
in $K$, serves only to normalize the group volumes to unity.

The character expansion of the Boltzmann weight is the finite sum
\bea
\exp[S(U)] = \sum_{n=0}^{P-1} \; b_{n} \; U^{n} \;\; ,
\eea
where the $P$ coefficients $\{b_{n}\}$ can be considered to be the
parameters of the theory. Indeed, one can invert this relation to obtain,
\bea
b_{n} = \frac{1}{P}\, \sum_{U\in Z_{P}} \; U^{-n}\, \exp[S(U)] \;\; .
\eea
It is usually required that the Boltzmann weight be insensitive to the
orientations  of the holonomies, $\exp[S(U)] = \exp[S(U^{-1})]$,
and this translates into the condition $b_{P-n} = b_{n}$. Without
this restriction, one must specify which orientations are being used in
(\ref{z}). Our analysis does not require this assumption, but
we will see that some formula simplify if it holds.

  Let us introduce an integer $n_{\Delta}\in \{ 0,...,P-1 \}$
for each 2-simplex $\Delta$, so ${\cal Z}_{d}$ becomes,
\be
P^{-N_{1}}\; \sum_{\{ U_{ij} \} }
\; \prod_{\Delta \in K^{(2)}} \;
\sum_{\{ n_{\Delta} \}} \; b_{n_{\Delta}} \; U_{\Delta}^{n_{\Delta}}
\;\; .
\ee
The collection of the $n_{\Delta}$ for all 2-simplices in $K$ may be
viewed as a 2-chain, which we can represent explicitly as,
\bea
n = \sum_{\Delta\in K^{(2)}} \; n_{\Delta} \, \Delta  \;\; .
\eea

Now rearrange the order of factors, the idea being to collect all
terms proportional to each link variable $U_{ij}$; we have
\bea
{\cal Z}_{d}= P^{-N_{1}}\;\sum_{\{ n_{\Delta} \}}\;
\prod_{\Delta \in K^{(2)}}
b_{n_{\Delta}} \; \prod_{ [i,j]\in K^{(1)} } \;
( \sum_{ U_{ij} } \; (
\prod_{\Delta \supset [i,j]}\; U_{ij}^{\varepsilon([i,j],\Delta)\,
n_{\Delta}} ))\;\; ,
\eea
where the last product in this equation is over all 2-simplices which
contain the specified link $[i,j]$. The factor $\varepsilon([i,j],\Delta)$
of $\pm 1$ explicitly records whether $[i,j]$ occurs in $\Delta$ with
positive or negative orientation.
Using the representation of a mod-$P$ delta function,
\be
\sum_{U\in Z_{P}} \; U^{n} = P \; \delta(n) \;\; ,
\ee
one obtains,
\bea
{\cal Z}_{d} = \sum_{\{ n_{\Delta} \}}\; \prod_{\Delta \in K^{(2)}}
b_{n_{\Delta}} \; \prod_{  [i,j]\in K^{(1)} } \;
 \delta(\sum_{\Delta \supset [i,j]}\; \varepsilon([i,j],\Delta)\,
n_{\Delta})\;\; .    \label{zdelta}
\eea
Notice that the sum in the delta function is over all 2-simplices
which contain $[i,j]$. Moreover, it is precisely the condition
that $n$ be a $2$-cycle ($\partial\, n=0$). The partition function is
then simply,
\bea
{\cal Z}_{d} = \sum_{ n \in Z_{2}(K,Z_{P}) }\;
\prod_{\Delta \in K^{(2)}} \, b_{n_{\Delta}}  \;\; .\label{z2}
\eea

Now, we can decompose the $2$-cycles in the following way,
\bea
Z_{2}(K,Z_{P})  = H_{2}(K,Z_{P}) \oplus \partial_{3} \, C_{3}(K,Z_{P})
\;\; , \label{cyc}
\eea
so the first part of the sum represents those $2$-cycles which are
nontrivial, and the second those which are trivial in the sense that
they are boundaries of $3$-chains. In this ``dual'' picture, we
see that the new dynamical variables are the $3$-chains, together
with a finite number of topological modes in correspondence with
$H_{2}(K,Z_{P})$. However, in summing over all of $C_{3}(K,Z_{P})$, we
would generically
overcount the number of independent $2$-cycles $n$ since $\partial_{3}$
may have a kernel. To simplify the computation of this kernel,
we restrict $P$ to
be a prime number so that $Z_{P}$ is an algebraic field and all
the chain and cycle groups are just vector spaces over $Z_{P}$.

   By definition of the boundary map we have
\bea
\partial_{m}: C_{m}(K,Z_{P}) \rightarrow C_{m-1}(K,Z_{P}) \;\; .
\eea
Taken together with the definition of homology, the two relations
\bea
dim\; Im(\partial_{m}) &=& dim\; B_{m-1} =
dim\; C_{m} - dim\; Z_{m}\;\; , \\
dim\; H_{m} &=& dim\;Z_{m} - dim\; B_{m}   \non
\eea
hold, and we can solve the recursion formula
\bea
dim\; Z_{m} = dim\;C_{m} + dim\;H_{m-1} - dim\; Z_{m-1}
\eea
which has the boundary condition $dim\; Z_{0} = dim\; C_{0} = N_{0}$.
For the situation at hand, we have that
\bea
dim\; ker(\partial_{3}) = N_{3}-N_{2}+N_{1}-N_{0} + h_{2} - h_{1} + h_{0}
\;\; ,   \label{knl}
\eea
where $N_{m} = dim\; C_{m}$ is the number of $m$-simplices in $K$,
and $h_{m} = dim\; H_{m}(K,Z_{P})$. For a connected complex $K$, which
we always assume, $h_{0} = 1$. Taking account of this kernel, we
have
\bea
{\cal Z}_{d} = P^{- dim\, Z_{3}(K,Z_{P})}\;
\sum_{ B \in H_{2}(K,Z_{P}) }\; \sum_{ C\in C_{3}(K,Z_{P}) }\;
\prod_{\Delta \in K^{(2)}} \, b_{(B+ \partial C)_{\Delta}}  \;\; ,\label{gz}
\eea
with $dim\; Z_{3}$ given explicitly by (\ref{knl}).

   The fact that $\partial_{3}$ has a kernel means that theory as formulated
in (\ref{gz}) has some gauge invariance; the number of gauge degrees of
freedom being precisely equal to the dimension of this kernel. This amount
of gauge redundancy is, however, far less than in the original link
based formulation.

In the dual theory, we see that the new Boltzmann weight is just proportional
to $b_{(B+\partial_{3} C)}$, with the variables $B$ and $C$ taking their
values in the additive group $Z_{P} = \{ 0,...,P-1 \}$.
We can easily revert to multiplicative notation if we wish,
\be
V_{[i,j,k,l]} = \exp[ \frac{2 \pi i}{P} \; C_{[i,j,k,l]}]
\;\;\; , \;\;\; W_{\Delta} = \exp[\frac{2 \pi i}{P} \;
B_{\Delta}] \;\; ,
\ee
and the dual action is a function of the product of the $W$ and $V$
variables. The $V$ variables that will enter a term
$\hat{S}_{\Delta}$ in the dual action
will be all those which have $\Delta$ as a face.

   The analysis of $U(1)$ gauge theories is very similar to the $Z_{P}$
case. If we replace the unit volume $Z_{P}$ group integration measure,
\bea
\frac{1}{P}\; \sum_{U\in Z_{P}} \rightarrow
\frac{1}{2\pi}\; \int_{0}^{2\pi} \, d\theta\;\; ,
\eea
were $U_{jk}=e^{i\theta_{jk}}$ is now the link variable associated
to the link $[j,k]$, the partition function is defined to be,
\bea
{\cal Z}_{d}[U(1)] = (2\pi)^{-N_{1}}\;
\prod_{[i,j]\in K^{(1)}}\; \int_{0}^{2\pi}\,
d\theta_{ij} \;
\prod_{ \Delta \in K^{(2)} } \; \exp [ S(U_{\Delta}) ] \;\; .
\eea
A character expansion for the Boltzmann weight $\exp[S(U)]$
is, in this case,  nothing other than a Fourier series,
\bea
\sum_{n= -\infty}^{\infty}\; b_{n}\, U^{n}\;\; .
\eea
The analysis leading up to equation (\ref{z2}) applies here as well,
only with the integer coefficient group $Z$ replacing
$Z_{P}$, and we obtain
\bea
{\cal Z}_{d}[U(1)] = \sum_{ n \in Z_{2}(K,Z) }\;
\prod_{\Delta \in K^{(2)}} \, b_{n_{\Delta}}  \;\; .
\eea

   To go beyond this general expression, some attention to the issue
of gauge fixing is required. The decomposition in (\ref{cyc}) can of
course be used (with the coefficient group $Z$), but we cannot sum
over redundant gauge field copies (which are in correspondence with the
group of $3$-cycles) since each would introduce an infinite factor via
a mode sum over $Z$.

\section{$d=2$ Gauge theory}

   Gauge theory on Riemann surfaces is particularly simple, and even in
the nonabelian case, the partition function for lattice gauge theory
has been computed as a function of the genus \cite{AM,EW}. However,
that analysis depended on choosing a Boltzmann weight which was
subdivision invariant; our analysis of the abelian case will make
no such restriction.

   Equation (\ref{gz}) for the partition function of $Z_{P}$ lattice
gauge theory is completely general. For $d=2$, there are no $3$-chains,
so $C_{3} = 0$ and $dim\, Z_{3} = 0$; our partition function then
reduces to a sum over $H_{2}(K,Z_{P})$,
\bea
{\cal Z}_{2} =
\sum_{ B \in H_{2}(K,Z_{P}) }\;
\prod_{\Delta \in K^{(2)}} \, b_{B_{\Delta}} \;\; .
\eea
For an oriented manifold, $dim\;H_{2}(K,Z_{P}) = 1$, and we have just
one mode sum to perform. The generator of $H_{2}$ is easily deduced
from the definition of $K$, and $B$ is then just a multiple of that
generator. In the four vertex complex of the $2$-sphere, for example,
the generator of $H_{2}$ is given by
\bea
[1,2,3] - [0,2,3] + [0,1,3] - [0,1,2]\;\; .
\eea
In fact, the fundamental class of any manifold
is such a sum of $d$-simplices with signs. In our partition function we
are summing over multiples $k$ of this generator so each $b_{B_{\Delta}}$
factor is then either $b_{k}$ or $b_{-k} = b_{P-k}$. When the action
$S$ is independent of the orientation of the holonomy (so $b_{k}=b_{P-k}$),
as is usually assumed, the partition function
for $Z_{P}$ gauge theory (P need not be
a prime number here) on an orientable surface reduces to
\bea
{\cal Z}_{2} = \sum_{k=0}^{P-1} \; b_{k}^{N_{2}} \;\; . \label{z2d}
\eea
It is independent of the genus and only depends on the
number of $2$-simplices in the triangulation. As such, one can
see which Boltzmann weight factors have a continuum limit
by examining the $N_{2} \rightarrow \infty$ limit of (\ref{z2d}).

   There is no essential difference in the analysis of the $U(1)$
case since $Z_{3}(K,Z)=0$, and we have,
\bea
{\cal Z}_{2}[U(1)] = \sum_{k=-\infty}^{\infty} \; b_{k}^{N_{2}}
\;\; .
\eea

\section{$d=3$ Gauge Theory}

In three dimensions, the situation is nontrivial but still very
regular, owing to the fact that each $2$-simplex in $K$ is the
face of precisely two distinct $3$-simplices. This gives rise to
an action which is a function of only two dynamical variables for
each $2$-simplex. For a closed, oriented $3$-manifold, the Euler
characteristic
\bea
N_{3} - N_{2} + N_{1} - N_{0}
\eea
vanishes and $dim\, Z_{3}$ given in (\ref{knl}) reduces to
$h_{2}-h_{1}+1 = h_{3} = 1$. The last equality follows from the
universal coefficient and duality theorems \cite{JM}. Alternatively,
we can view this as a specific confirmation of our general formula
(\ref{knl}), since we know that in three dimensions $Z_{3}(K,Z_{P})
= H_{3}(K,Z_{P})$ (as $B_{3}(K,Z_{P})=0$).

Let us illustrate the formula in detail for the complex (\ref{cs3}) of
the $3$-sphere. Since $H_{2}(S^{3},Z_{P})=0$, we have only to sum over
all $3$-chains in computing the partition function.
An arbitrary $3$-chain $C$ can be written as
\bea
C = c_{0}\, [1,2,3,4] - c_{1}\, [0,2,3,4] + c_{2}\, [0,1,3,4]
- c_{3}\, [0,1,2,4] + c_{4}\, [0,1,2,3] \;\; ,\label{C}
\eea
where each $c_{a}$ is an element of $\{ 0,...,P-1 \}$ and  arithmetic
is all modulo $P$. We need not have included the minus signs in (\ref{C}),
but they make the following expressions more symmetrical. The partition
function on this complex (assuming for simplicity that the action is
independent of the orientation of the holonomy) then reduces to,
\bea
\frac{1}{P}\; \sum_{c_{0},...,c_{4}}\;
b_{c_{3}-c_{4}} \,
b_{c_{2}-c_{4}}\,
b_{c_{2}-c_{3}}\,
b_{c_{1}-c_{4}}\,
b_{c_{1}-c_{3}}\,
b_{c_{1}-c_{2}} \,
b_{c_{0}-c_{4}} \,
b_{c_{0}-c_{3}} \,
b_{c_{0}-c_{2}}\,
b_{c_{0}-c_{1}} \;\; .
\eea

As we explained above, the kernel of $\partial_{3}$ is one
dimensional and we have the gauge freedom to arbitrarily set any one of the
$c_{a}$ variables to zero. This is substantially less gauge freedom
than we had in the original link based formulation.

  Equation (\ref{gz}) has been checked numerically on a small cubic
lattice with a specific action and gauge group $Z_{2}$. Since the
$3$-torus $T^{3}$ has $H_{2}(T^{3},Z_{2}) = Z_{2}\oplus Z_{2}\oplus
Z_{2}$, we have a total of $2^{3}=8$ topological modes, and each of the
three generators is represented by an embedded $2$-torus. For a
$3 \times 3 \times 3$ lattice, the dual partition function can be
summed exactly and compared to a Monte Carlo approximation to the
original gauge theory formulation (this has many more modes and
cannot be summed exactly). This numerical check reveals that the
individual topological sectors of the dual theory are generally
distinct, and some make a negative contribution to the partition
function.

The $U(1)$ gauge theory on a closed, oriented 3-manifold,
parallels the above, only we are forced to gauge fix the dual theory
to get a meaningful result. One has,
\bea
{\cal Z}_{3}[U(1)] = \sum_{B\in H_{2}(K,Z)} \; \sum_{C\in C'_{3}(K,Z)} \;
\prod_{\Delta\in K^{(2)}}\; b_{ (B+\partial C)_{\Delta}}
\;\; ,
\eea
where $C'_{3}(K,Z)$ is the gauge fixed set of all $3$-chains. The later
differs from $C_{3}(K,Z)$ only in having set a single chosen component
to an arbitrary value in $Z$.

\section{Concluding Remarks}

   The appearance of homology
modes in duality transformations is generic, and not specific to the abelian
gauge theory that we treated in this paper. These modes are analogous to
the cohomology
modes which are present in dual cell complex formulations \cite{RAK}.

   Given that duality exchanges strongly and weakly coupled theories, the
need to gauge fix the dual of $U(1)$ lattice gauge theory
is not really surprising.
This parallels the continuum situation where weak coupling is treated
perturbatively, and gauge fixing is required.

{\Large\bf Acknowledgments }\\
The work of S.S was partly supported by Forbairt Contract Sc/208/94.

\end{document}